# Dielectric resonator antenna for 3D uniform manipulation of NV centers in diamond

Polina Kapitanova[1,*], Vladimir V. Soshenko[2,*], Vadim V. Vorobyov[2,3,*], Dmitry Dobrykh[1] Stepan V. Bolshedvorskii[2,3|], Vadim N. Sorokin[2,4], and Alexey V. Akimov[2,4,5]

*[1]Department of Nanophotonics and Metamaterials, ITMO University, 197101 Saint Petersburg, Russia*

*[2]P.N. Lebedev Physical Institute, 119991 Moscow, Russia*

*[3]Moscow Institute of Physics and Technology, 141700 Moscow Region, Russia*

*[4]Russian Quantum Center, 143025 Moscow Region, Russia*

*[5]Texas A&M University, College Station, TX 77843, USA*

Ensembles of nitrogen-vacancy (NV) color centers in diamond hold promise of ultraprecise magnetometery competing with SQUID detectors. By utilizing advantages of dielectric materials such as very low losses for electromagnetic field and therefore possibility to create high quality factor resonators with strong concentration of the field in it we implemented a dielectric resonator antenna for coherent manipulation of large ensemble of NV centers in diamond. We reached average Rabi frequency of 10 MHz in the volume of 7 cubic millimeters with standard deviation less than 1% at moderate pump power. This result passes the way to improve sensitivity of cutting edge NV based magnetometers by two orders of magnitude practically reaching SQUID level of sensitivity.

## I. INTRODUCTION

Color centers in diamond have been the focus of significant interest over the last decade. In particular, many metrological applications have been found, including ultra-precise magnetometers [1–4], biocompatible thermometry [5,6], electron and nuclear magneto resonance imaging [7–9], electric field sensors [10,11], and strain sensors [12]. Many of these applications rely on large ensembles of NV centers in a single diamond plate, which make the detector sensitive. For example, with diamond bulk crystals, sensitivities at subpicotesla levels both in AC and DC regimes have been demonstrated already [3,4].

Sensor and metrology applications for situations when high sensitivity is concerned require efficient driving of the NV centers' electronic spins within large volume. This places demand on the microwave antenna to produce a uniform and strong microwave (MW) magnetic field over a relatively large (few cubical millimeter) volume. A non-uniform MW field will lead to different Rabi frequencies for different regions of the sample, resulting in different dynamics for different parts of the ensemble. This could be partially overcome by implementation of sequences robust to variation of parameters [13], but ultimately, uniformity of the MW magnetic field limits the number of NV centers one could use. Recently, a double split-ring resonator (SRR) antenna implemented as a printed circuit board (PCB) was proposed for planar diamond substrates [14]; however, its MW field decays dramatically in perpendicular direction from the surface of the PCB, providing only a 2-dimensionally uniform field. Here, a dielectric resonator antenna (DRA) utilizing a high-permittivity, low-loss dielectric resonator and providing a 3D-uniform MW field in sufficient volume to efficiently excite NV centers in whole, commercially available samples, is proposed. This antenna design, along with recent techniques for optical, large volume excitation [4], paves the way for room temperature femtotesla range magnetometers, electric field, and strain sensors.

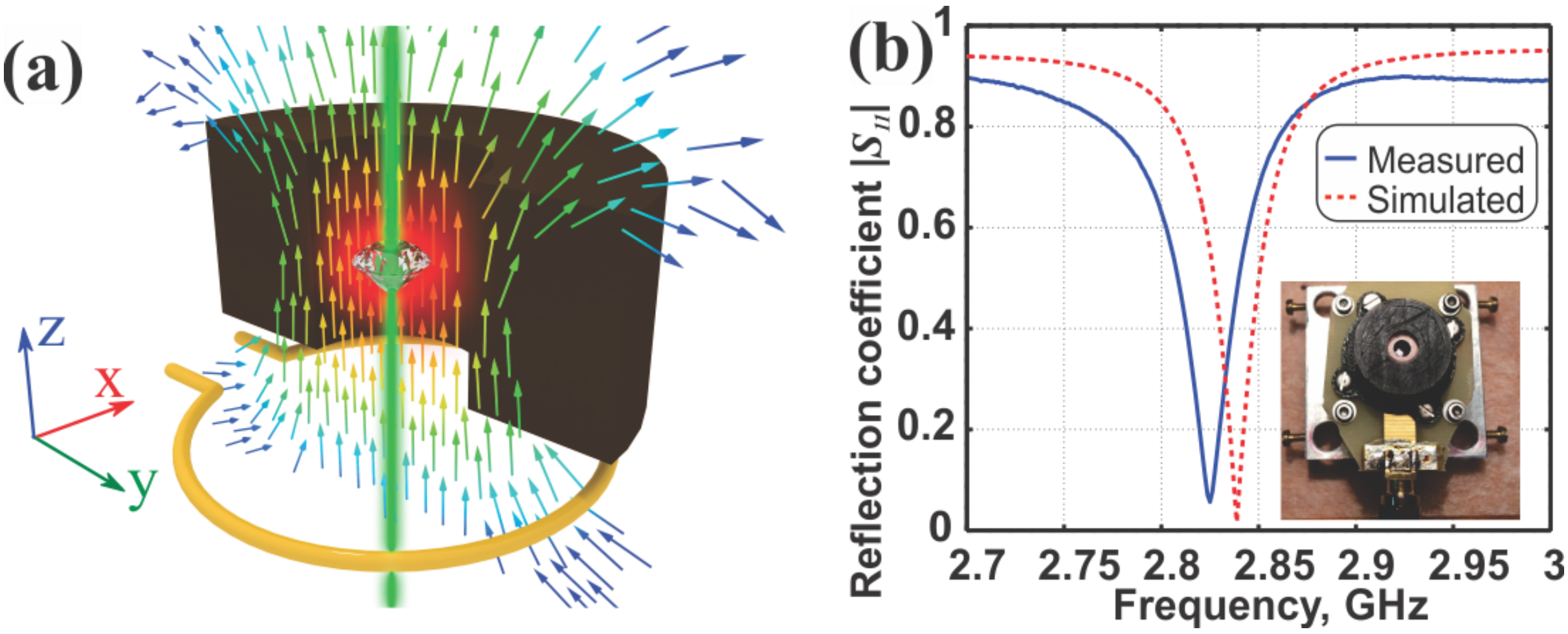

*FIG. 1. (a) Design of the DRA. (b) Simulated and measured reflection coefficient of the DRA. The inset demonstrates the DRA prototype photograph. distribution have been performed in CST Microwave Studio for the input power level of 0.5 W.*

## II. DIELECTRIC RESONATOR ANTENNA.

The idea behind implementation of the DRA is to on one side use macroscopically large but uniform field mode of dielectric resonator [15] and on the other side, use the fact, that the resonator can accumulate quite a lot of field within. Luckily high Q-factor, compact size, low weight dielectric resonators become available this days and could be produced at low cost [15].

The design of the DRA for efficient magnetic coupling to NV centers in diamond is based on excitation of the TE01δ mode of a hollow cylindrical dielectric resonator using a small coupling loop, as shown in FIG. 1(a) (see *Appendix A* for details). Thus, the MW magnetic field is concentrated inside the cylindrical resonator and oscillating along the $z$ axis. In this case, the optical sample can be placed in the bore of the resonator where the MW magnetic field is maximal.

The simulated and measured reflection coefficients of the DRA are shown in FIG. 1(b). The inset demonstrates the DRA prototype photograph. The simulated reflection coefficient of the DRA has a characteristic minimum at the frequency of 2.84 GHz that corresponds to the maximum of the power that couples to the resonator from a MW source. The measured reflection coefficient has a minimum at the frequency of 2.837 GHz. The slight red shift of the frequency can be explained by the holder influence, which was not taken into account during the numerical simulations (see *Supplemental Material* for details). The unloaded Q-factor of the dielectric resonator is 3000 at the frequency of 2.84 GHz. The loaded Q-factors extracted from the simulated and measured reflection coefficients by the standard technique at the 0.5 level are 150 and 106, respectively [16].

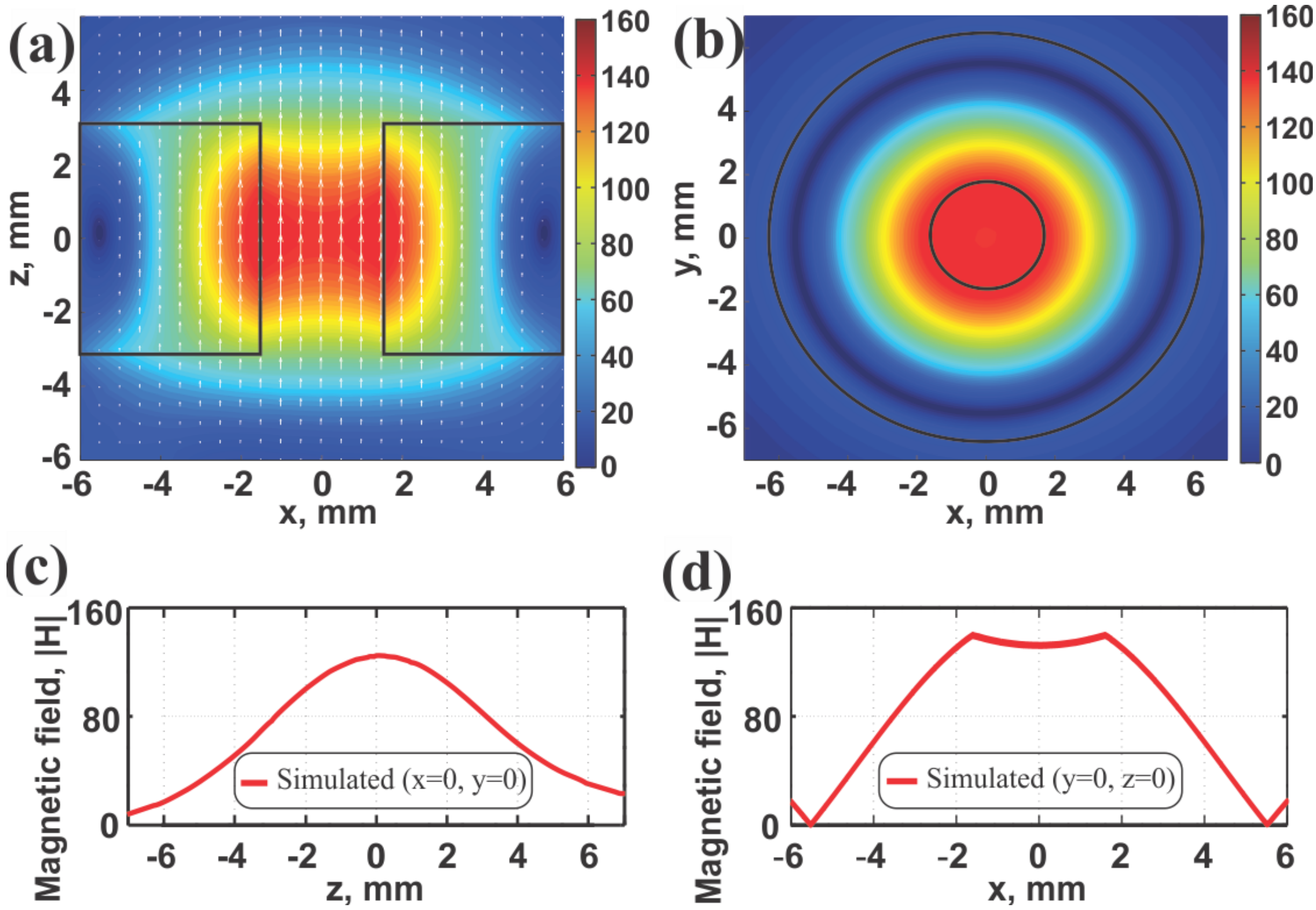

*FIG. 2. (a) Simulated magnitude of the MW magnetic field distribution in $xz$ plane of the DRA at the frequency of 2.84 GHz. (b) Same for the $xy$ plane. (c) and (d) MW magnetic field magnitude profile as a function of coordinate simulated at the frequency of 2.84 GHz. The simulations of the reflection coefficient and the magnetic field distribution have been performed in CST Microwave Studio for the input power level of 0.5 W.*

The simulated magnitudes of the MW magnetic field distributions at the resonance frequency are shown in FIG. 2(a) and (b) in the $xz$ plane and $yz$ plane of the dielectric resonator, respectively. One can see that the magnetic field amplitude along the $z$ axis is uniform and mostly concentrated in the entire bore of the dielectric resonator. By setting the coordinates to $x = 0$, $y = 0$, $z = 0$ corresponding to the dielectric resonator center, and plotting the magnetic field distribution as a function of the coordinates we obtained that the field behavior inside the DRA bore along the $z$ axis has a local maximum in the center of the bore (see FIG. 2(c)). In radial direction, the field has a minimum at the axis of the DRA due to the presence of the bore (see FIG. 2(d)). Still, for the 1.6 mm bore radius, the field variation is less than 17% in $z$ direction and 6% in radial direction.

The presence of the bore is very accommodating to an optical sample. Moreover, as soon as the uniform MW magnetic field distribution is obtained in large volume, the optical sample can be shaped differently and of relatively large size.

# III. RESULTS AND DISCUSSION

The practically important value that defines an NV-based sensor's performance is Rabi frequency achieved in the MW field created by the DRA. To test the uniformity of the Rabi frequency for selected orientation of an NV center, a bulk diamond sample with a relatively high concentration of NV centers has been placed in the middle of the DRA bore and experimentally studied using a home-built scanning microscope as depicted in FIG. 3(a). It provides enough distance and resolution to perform scanning of the Rabi frequency of NV centers inside the DRA (see *Appendix B* for details).

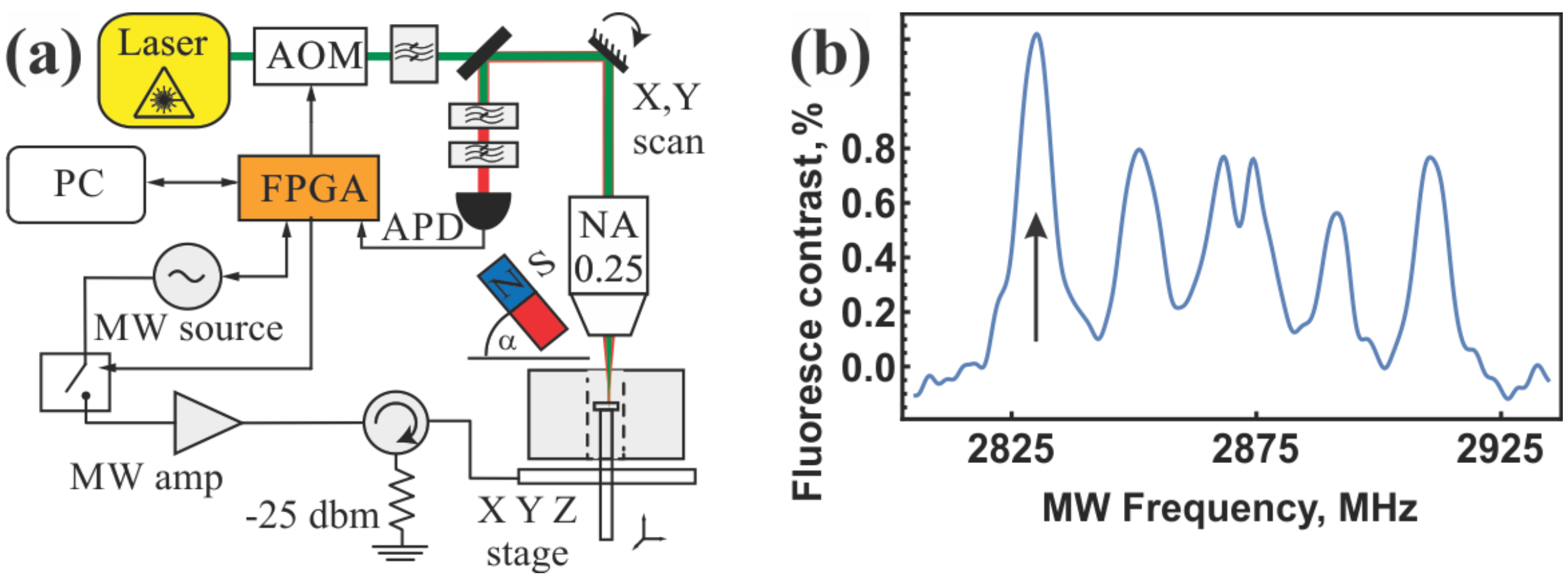

*FIG. 3. (a) Scanning microscope setup with a 0.25 NA objective lens for selective addressing of NV centers sub-ensemble. (b) ODMR of a single-orientation resonance, selected by the MW magnetic field (continuous excitation). The arrow indicates the resonance selected for further measurements.*

By scanning the direction of the constant in a magnitude bias magnetic field using a continuous excitation scheme (see *Supplemental Material*) we were able to monitor the frequencies of all 4 orientations and shift the resonance line corresponding to $\bar{1}11$ orientation to the minima of the DRA reflection coefficient (30 MHz full width half-maximum, see FIG. 1(b)) having the other orientations detuned (see FIG. 3(b)).

The uniformity of the MW magnetic field is confirmed through an experimental study of the Rabi frequency of NV centers with different positions of the diamond sample inside the DRA bore (see *Supplemental Material*). FIG. 4(a) demonstrates the measured distribution of the Rabi frequency in the radial direction in comparison to the results predicted by the numerical simulations. The simulated and measured Rabi frequencies are different by about 2% in amplitude, while the distribution is almost exactly the same in both cases. The discrepancy most likely is caused by losses in the connector feeding the DRA and the influence of the surrounding optical equipment. One of the key advantages of this design is the long depth of the MW magnetic field along the $z$ direction. The measured Rabi frequency as a function of the diamond position along the $z$ direction is presented in FIG. 4(b) in comparison to the Rabi frequency of the planar SRR design [14]. The Rabi frequency of the planar SRR design clearly decreases drastically with the increasing of the distance along the $z$ direction, while the DRA still provides the high values of the Rabi frequency. Nevertheless, a practically important parameter affecting the sensitivity of magnetometers and diamond-based sensors is the volume of excitation possible with the DRA design in uniform manner. FIG. 4C provides insight on the MW magnetic field uniformity inside a cylindrical ensemble of NV centers with the diameter of $D$ and height of $H$. The color map represents the fitted measured data on average over volume Rabi frequency of the ensemble of NV centers (see *Supplemental Material* for details). It is more convenient to think of field quality in terms of variation of the field in a given volume. The relative average Rabi frequency inhomogeneity is presented in FIG. 4C as isolines. One can see from the figure that in the cylindrical ensemble of NV centers with dimensions of $1.7\ \text{mm} \times \varnothing 2.4\ \text{mm}$ the inhomogeneity of the average Rabi frequency was found to be less than 1%. With this volume, the sensitivity of an ultra-precise, diamond-based magnetometer could be improved by a factor of 90 from $0.9\,\text{pT}/\sqrt{\text{Hz}}$ to $\sim 10\,\text{fT}/\sqrt{\text{Hz}}$ due to the increase in number of NV centers (see Appendix D) [3]. Further minimization of the field inhomogeneity could be achieved by selecting the optimal size of the ensemble of color centers, which keep minimal the non-uniformity of the field. This optimal size is represented by the dashed line in FIG. 4(c). FIG. 4(d) is directly refers to that curve and shows the minimum possible non-uniformity of the MW magnetic field for a given volume of NV ensemble. A comparison with the average Rabi frequency inhomogeneity seen in FIG. 4(d) shows the advantage of the proposed DRA over the SRR design [14].

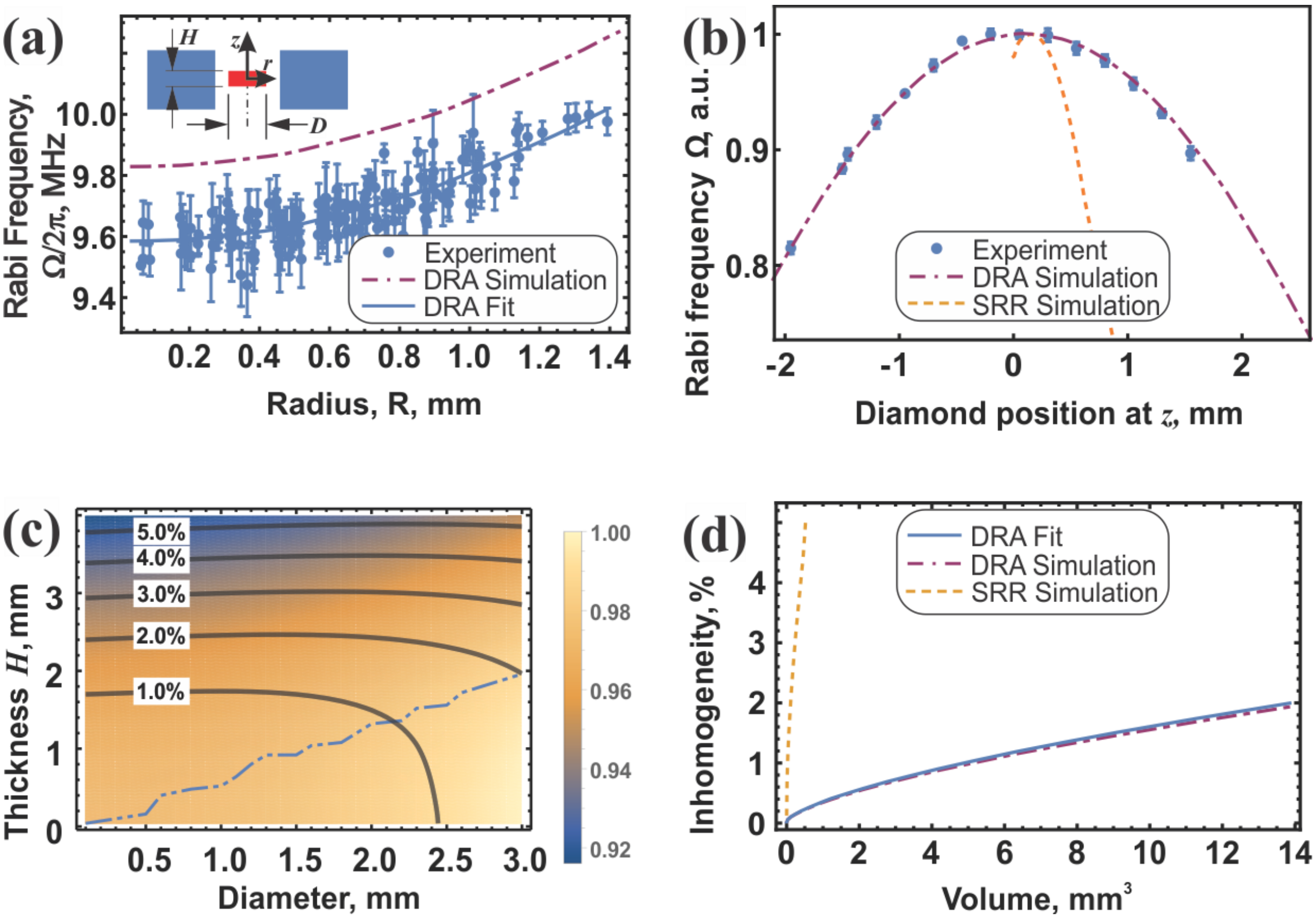

*FIG. 4. (a) Experimentally measured (dots), fitted (solid line) and simulated (dash-dot line) Rabi frequency distribution in the radial direction. The inset illustrates the geometry of the experiment. The red rectangle represents the diamond plate, and blue rectangles, the DRA. (b) Normalized Rabi frequency dependency versus the diamond plate position along the $z$ axis (see inset in (a)). The orange dashed line represents simulation of the Rabi frequency in case of the SRR [14]; the magenta dash-dot line corresponds to the simulated one in the case of the DRA; and the dots,*

*to the experimental data taken using the DRA. The maximum value of the Rabi frequency measured with the DRA is $\Omega = 2\pi \cdot 10$ MHz at 5.2 W feeding power. (c) Measured uniformity of the Rabi frequency against diamond dimensions. The color map corresponds to the normalized average over volume bounded by current point Rabi frequency of ensemble. Contours indicate the level of inhomogeneity (%) within the volume bounded by each contour. The dash-double dot blue line shows the optimal dimensions of the diamond sample with minimum field inhomogeneity for a certain volume (see the inset to (a)). (d) The inhomogeneity of the MW magnetic field versus volume of the optimal diamond sample (where the solid blue line represents fitted experimental data; the dash-dot magenta line, simulation for DRA; and the orange dashed line, simulation for SRR* [14]*).*

Finally, to demonstrate the DRA's performance in a realistic device, the fluorescence contrast of the large ensemble of NV centers was measured using the volume collection setup depicted in FIG. 5(a) (see *Appendix C* for details). To this end, the efficiency of the DRA manifested as magnitude of ODMR contrast or Rabi oscillations amplitude. FIG. 5(b) demonstrates the measured fluorescence contrast of 6% with the Rabi frequency as high as 8 MHz. It should be stressed that this fluorescence contrast value is nearly the maximum possible for a single orientation of NV center. The slightly lower value of the Rabi frequency in this case, compared to the measured one in the scanning setup, is due to the fact that, due to geometrical limitations, the diamond plate was placed at a distance of $z = -2$ mm from the dielectric resonator center (see FIG. 5(a)). In this plane, the MW magnetic field amplitude reaches 100 A/m, while the maximum field of 132 A/m created by the DRA is in the center of the dielectric resonator when $z = 0$ (FIG. 2(a)). In addition, the sample had a $\bar{1}11$ cut, leading to an angle of about 109.5 degrees between the MW magnetic field and spin quantization axis. This slightly (about 6%) reduces the achieved Rabi frequency in both experiments.

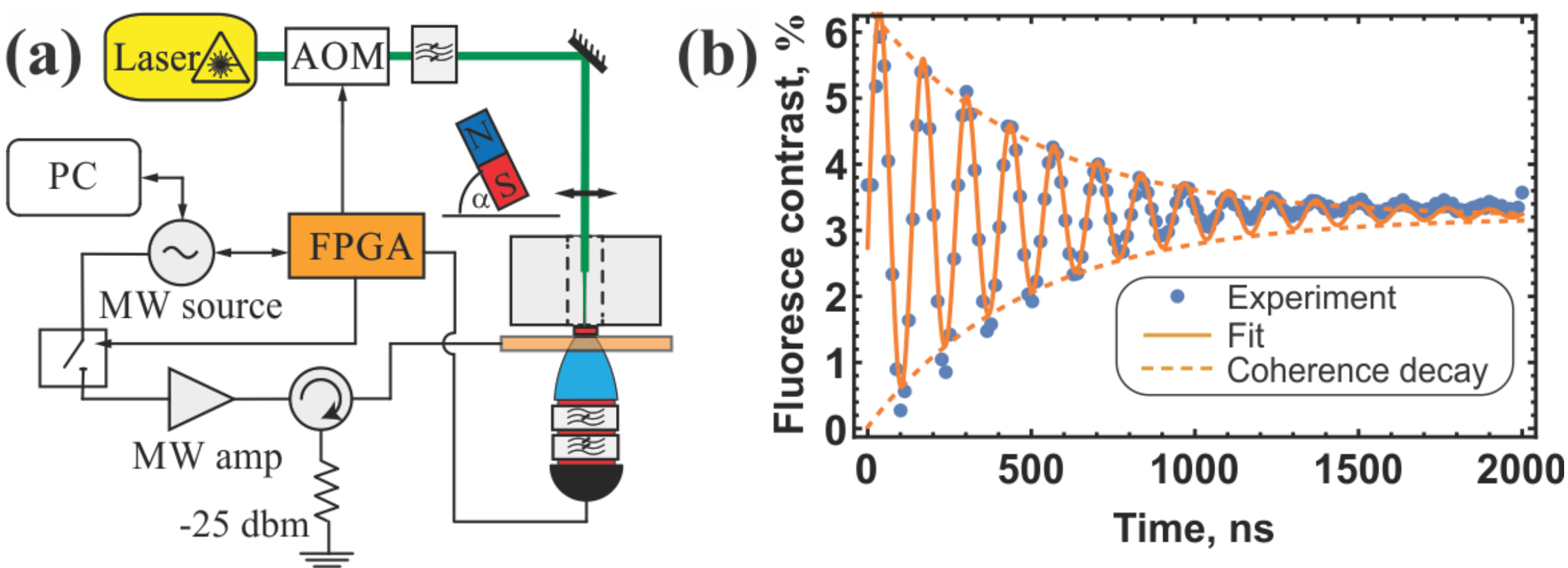

*FIG. 5. (a) Experimental setup for bulk collection from large ensemble. (b) Measured Rabi oscillations in ensemble of NV centers. 8 MHz Rabi frequency is reached at 5.2 W feeding power.*

The values reached in the Rabi frequency are not limitations of the DRA design. It is believed that further optimization of the DRA designs will provide higher efficiencies. As an alternative to the obvious option of the DRA supplying more power, one could use other types of microwave ceramic characterized by higher permittivity (up to 1000) and lower value of dielectric loss ( $\tan(\delta) < 10^{-4}$ ). Further optimization of the DRA feeding type also would provide advantages. Moreover, excitation of the higher-order modes of a dielectric resonator would produce a different orientation of MW magnetic field or move the field toward the side of the dielectric resonator to operate with other orientation of NV centers in diamond.

## IV. CONCLUSION

The dielectric resonator antenna for homogeneous control of spatially large spin ensembles has been designed, fabricated, and experimentally studied. The measured standard deviation of the average Rabi frequency at the volume of 7 cubic millimeters (dimensions 1.7 mm$\times\varnothing$2.4 mm ) was demonstrated to be less than 1%, while the magnitude of the MW magnetic field was enough to reach 10 MHz average Rabi frequency at 5.2 W feeding power. Application of the proposed antenna in modern, NV-based, room temperature AC magnetometers would allow two-fold enhancement in sensitivity. Even higher efficiency of the DRA without loss of field uniformity is still possible via utilization of higher finesse resonators or higher-order modes of resonators. This work paves the way for building ultra-precise sensors, based on dense spin ensembles in solid state, such as NV centers in diamond.

## ACKNOWLEDGMENT

The authors are grateful to I. V. Shadrivov for useful discussions and to E.A. Nenasheva for providing the microwave ceramic resonator samples. This work was supported by the Russian Science Foundation (Grant #16-19-10367). P. K. acknowledges a scholarship from the President of the Russian Federation.

## APPENDIX A: DIELECTRIC RESONATOR ANTENNA

The dielectric resonator itself is made of $BaLn_2Ti_4O_{12}$ microwave ceramic [17], which is characterized by the permittivity of $\varepsilon = 80$ and loss tangent $\tan(\delta) = 0.0003$ (measured at the frequency of 2.84 GHz). The DRA design has been optimized using a frequency solver of CST Microwave Studio (see *Supplemental Material* for details). The dielectric resonator dimensions are: an outer diameter of $D_{out} = 12.5$mm , height of $L = 6$mm and bore diameter of $D_{in} = 3.2$ mm .

Modes excited within a cylindrical DRA are classified into three distinct groups: TE, TM, and HEM (hybrid) modes. In antenna design, a cylindrical dielectric resonator usually is mounted on a ground plane and fed by a coaxial probe, slot coupling, or microstrip line [18–20]. In this way, the TM or HEM mode is excited in the dielectric resonator and provides electric, dipolar-like radiation characteristics in the far field. As soon as one addresses the NV centers' excitation, one needs to provide strong magnetic field; in other words, one needs to excite the TE mode of a dielectric resonator [21–24]. In case of applying of external magnetic field , one must excite the $TE_{01\delta}$ mode of the dielectric resonator which is characterized by strong magnetic field oscillating along the resonator axis. This is done using a symmetric microstrip loop placed next to the cylindrical dielectric resonator.

## APPENDIX B: SCANNING SETUP

The setup (FIG. 3 (a)) was mostly the same as in reference [25] except for the objective. A 10X Olympus Plan Achromat Objective, with numerical aperture of 0.25 NA and objective lens working distance of 10.6 mm, was used. The bias magnetic field was created using a permanent magnet, which was placed on a controllable positioner. Microwave pulses with maximum output power of 5.2 W were generated using computer controlled field-programmable gate array board (see *Supplemental Material* for microwave pulses details).

As a sample we used 1.2 x 1.2 x 0.3 mm diamond plate with NV concentration about 13 ppm; produced by the LLC Velman company.

## APPENDIX C: ENSEMBLE SETUP

The NV ensemble emission was collected using a compound parabolic concentrator (Edmund Optics CPC) which was glued directly to the sample (FIG. 5(a)). This scheme allows collection of more than 60% of NVs fluorescence [3] The NV ensemble was optically pumped with a 532 nm laser of 1 W (Verdi Coherent V8). Green radiation was focused onto a diamond surface with a 100 mm focus lens. The fluorescence outgoing from the parabolic concentrator was filtered with a notch 532 nm and longpass 600 nm filters (Semrock), and focused with a 25 mm antireflection coated lens onto a silicon pre-amplified photodiode (Thorlabs PDA 100A).

## APPENDIX D: ESTIMATION OF MAGNITOMETER SENSITIVITY

According to the Reference [3] the sensitivity of a magnetometer equal to $0.9\, pT/\sqrt{Hz}$ can be obtained for the following parameters:

Effective volume of NVs: $8.5\times10^{-4}$ mm$^3$,

NV concentration: 0.9 ppm,

Estimated number of NVs: $1.4\cdot10^{11}$

If we consider the application of the proposed DRA we increase the efficient volume of NVs up to 7 mm$^3$. It is approximately 8200 times larger than in [3], hence, the signal to noise ratio and sensitivity of AC magnetometry can be improved approximately 90 times. Thus, the sensitivity one could obtain is $10\, fT/\sqrt{Hz}$ instead of $0.9\, pT/\sqrt{Hz}$.